# A disclosure of the role of frequency on the plasma electrolytic oxidation of AZ31 magnesium alloy in aluminate-tungstate electrolyte

Wenbin Tu, Qiushui Feng, Yingliang Cheng [*]

*College of Materials Science and Engineering, Hunan University, Changsha, 410082, China*

**Abstract:** Plasma electrolytic oxidation (PEO) was carried out to form black ceramic coatings on AZ31 magnesium alloy in aluminate-tungstate electrolyte. Influences of current frequency (such as 100, 1000 and 2000 Hz) on properties of the PEO coatings have been explored. It was found that increment of frequency led to coatings with higher thickness, higher roughness and darker appearance. In order to explore the coating formation mechanism at different frequencies, the micro-discharges at different frequencies were investigated by real-time imaging method. Microstructure of coatings was characterized by scanning electron microscopy (SEM) assisted with energy dispersive X-ray spectroscopy (EDS). Coatings formed at 100 Hz show the most compact microstructure while porosity of the coatings, both in the pore number and pore size, is greatly increased at higher frequencies. The coatings formed at higher frequencies show an increased atomic ratio of W/Al. The cancel of the negative pulse from the pulsed waveform can also lead to a higher W/Al ratio, and hence a darker coating. The higher porosity of the coatings was attributed to the stronger plasma discharges and gas evolution at higher frequencies. The stronger plasma discharge is induced by the fact that there is not sufficient time for the cooling down of discharge channels at higher frequencies.

**Keywords:** Plasma electrolytic oxidation, Magnesium alloy, Frequency, Real time imaging,

*Corresponding authors. Tel.: +86 731 88821727 ; Fax: +86 731 88823554
E-mail addresses: chengyingliang@hnu.edu.cn, deepblacksea@163.com(Y. Cheng).

Mechanism.

## 1. Introduction

As one of the lightest engineering metals, magnesium has attracted increasing attention. It has many excellent properties like high strength to weight ratio, good castability, good dimensional ductility and easy of recycling [1,2]. Those properties have made magnesium and its alloys widely applied in fields like automobile, aerospace, electro-communication, and biomedical application [3,4]. Whereas, some undesirable demerits of magnesium and its alloys such as high chemical activity and negative electrochemical potential weaken their corrosion and wear resistance, particularly in corrosive environment [5,6]. To cope with those two issues, many kinds of surface modification techniques have been explored by researchers such as electrodeposition, organic coatings, sol-gel technique, gas-phase deposition, and anodization; each has its own advantages and disadvantages [7-10]. Comparing with those methods, plasma electrolytic oxidation (PEO) is an inexpensive and environmentally friendly technique which has been broadly applied in surface treatment for magnesium, its alloys and other valve metals like aluminum and zirconium.

PEO treatment can fabricate thick ceramic layers on valve metals' surface to improve their serve performances. For instance, corrosion resistant PEO coatings were successfully formed via PEO technique combined with addition of nanoparticles into electrolyte or sealing treatment [11-14]; wear resistance of metal substrates was found to be effectively improved by fabricating PEO coatings on the metal surfaces [15-17]. Despite the fact that various PEO coatings could be formed, the properties of those coatings depend on a number of factors like sort of electrolyte [18,19], composition of substrate [20,21] and electric parameters [22,23].

Among those factors, electric parameters play significant role which should not be ignored during PEO treatment. Efforts have been made by many scientists on those parameters to optimize the properties of PEO coatings so as to broaden the applications of this technique and better understand the coating formation mechanism. Influences of applied voltage, current density and duty cycle on process of PEO treatment and properties of coatings have been broadly studied [22,23-26]. Despite much investigation on electric parameters of PEO treatment has been made, a systematic study concerning influence of current frequency on discharge behaviors during PEO treatment is less reported. Increment of current frequency was reported to result in a decreased pore size and less content of electrolyte species in the coating formed on aluminum alloy by Arunnellaiappan et al. [27]. A similar conclusion was drawn by Kaseem et al., [28] who reported that both the average size of micro pores and their porosity decreased with increasing current frequency, however, they pointed out that more species from electrolyte was incorporated into coatings at a higher frequency which was contrary to the phenomenon reported in [27]. On the other hand, Srinivasan et al. [29] found that lower frequency yielded thicker and rougher PEO coatings with more compact microstructure. Up to now, the influence of current frequency on PEO is still in controversy. The current frequency applied in most of those studies only covered a low range, such as from 50 to 1200 Hz, coatings formed at higher frequency like 2000 Hz were not compared. Furthermore, most of the reports were only focused on the changes in coating microstructure under different frequencies, and tried to relate the coating properties, such as corrosion resistance, to the corresponding changes in microstructure [28], seldom have they explored the underlying mechanism of the function of frequency that leads to these changes.

Apart from forming protective ceramic layers on metal substrate, PEO technique has also been used to synthesize coatings with diverse colors by adding some additives into electrolyte [30-32], or through oxidizing the elements in the substrate [33-35] which could generate colored oxides after PEO treatment. Nevertheless, it took several steps in those studies to alter coating color appearance, such as re-preparation of electrolyte; few reports have discussed the relationship between the coating color and the electric parameters like frequency or current regime. Meanwhile, most of those studies only qualitatively described the color of coatings instead of quantitatively representing the color difference through scientific data.

The application of a negative pulse (cathodic current) to the pulsed waveforms can also influence the PEO coating formation. Cathodic current was reported to cause soft sparking mode of PEO on aluminum by Rogov et al. [36]. Current mode without negative pulse was found to favor the formation of pancake surface morphology while bipolar converted the morphology to crater-like structure [37]. In the authors' previous work [38], negative pulse was found to facilitate the incorporation of more electrolyte species into coatings to form a homogenous ceramic layer on Zirlo alloy, but what influence will negative pulse pose on incorporation of ions in electrolyte containing two major species instead of only one species is not clear yet; on the other hand, now that most of colored PEO coatings are formed via incorporation of electrolyte species, whether or not can the color or brightness of coatings could be altered by electric parameters such as frequency and negative pulse is still undisclosed.

Consequently, in this work, PEO of AZ31 magnesium alloy was carried out in sodium-tungstate electrolyte at various frequencies to mainly study the underlying mechanism for the influence of current frequency on properties like thickness, roughness and microstructure of PEO coatings. In addition, the Commission Internationale d'Eclairage (CIE) L*a*b* coordinate and EDS analysis were applied to quantitatively represent color difference of the coatings at different electric parameters and understand the brightness alteration mechanism respectively.

## 2. Experimental

A rolled AZ31 magnesium alloy with a thickness of 6 mm was used as the substrate of PEO treatment. Prior to the treatment, the samples with a working area of 10 x 20 mm were grounded and polished with SiC abrasive paper (120-2000#), washed with acetone and distilled water. A stainless steel was used as the cathode while the substrates were connected to the positive pole of the power supply working as anode. Aqueous electrolyte of 10 g l$^{-1}$ NaAlO$_2$ + 10 g l$^{-1}$ Na$_2$WO$_4$ 2H$_2$O + 3 g l$^{-1}$ C$_6$H$_8$O$_7$ H$_2$O (citric acid) + 2 g l$^{-1}$ KOH was used for the PEO treatment. Constant current mode was employed with duty cycle being kept at 20%. Coatings were fabricated at frequencies of 100, 1000 and 2000 Hz with/without negative current. Details of experimental arrangement were largely the same as that in our previous paper [39,40].

The current waveforms under different frequencies were monitored using an oscilloscope (Tektronix TDS 1002C-SC). The surface and cross-sectional morphology of the coatings were examined by a field emission gun scanning electron microscopy (SEM, QUANTA 250, FEI,

USA) assisted by energy dispersive spectroscopy (EDS). The thickness of the coatings was determined by the average value of thickness measured at 12 different points on the coating surface using an eddy current thickness gauge (TT260, Time group, Beijing). The same method was applied to determine the surface roughness of coatings by a stylus profilometer (Mitutoyo SJ-210). To get a better understanding about difference of PEO process under various conditions, the volumes of gas released during PEO treatment under different frequencies were collected by liquid displacement method, as described in [41]. The volume of gas released was determined by the average value calculated after repeating three times for each condition so as to avoid the one-time occasionality in data-collection and ensure the validity of the tests. The real time appearance of plasma discharges during PEO was recorded by a commercial digital camera (Canon EOSD300) with an exposure time of 0.125 ms, which is the shortest exposure time of the camera. The influence of frequency on the distribution of the microdischarges on the sample surface was analyzed to obtain coating formation mechanisms. The same camera was also used to record the image of the appearance of the coatings.

The CIE (Commission Internationale d'Eclairage) L*a*b* color space (also known as CIELAB), which was designed to be perceptually uniform with respect to human color vision, is a model applied to express color as three numerical values, namely L*, a* and b*. It is used to quantitatively represent color difference in scientific researches [33,42,43]. In order to investigate the influences of current frequency and pulse regime on the brightness of the coatings, a color difference meter (CM-5, Conica Minilta Co., Japan) was used to measure the color of coated sample surfaces by applying the CIE L*a*b* values with a 10 ° standard observer and standard illuminant D56. The brightness of color was represented by L* value while a* and b* values indicated the location of color in the red-green and yellow-blue color coordinate. All the values of L*a*b* were the average values of three tested points so as to ensure the validity of the tests. Color difference (△E*) of ceramic coatings formed under different conditions was determined by applying the CIE color difference formula see as follows:

$$\Delta E* = \sqrt{(\Delta L*)^2 + (\Delta a*)^2 + (\Delta b*)^2}$$

## 3. Results

### 3.1 Cell potential--time responses, current waveforms and discharge appearance at different conditions

The cell potential-time responses for the PEO of magnesium alloy under pulsed bipolar and unipolar regimes with different frequencies are plotted in Fig. 1. As depicted in Fig. 1(a), the positive cell potential in the bipolar regime surges to high values at initial stage, reaching ~ 463, ~466 and ~467 V at 27, 26, 24 s for the samples under 100, 1000 and 2000 Hz, respectively. Afterwards, the cell potential rises with significantly reduced rates to the final potentials of ~534, ~568 and ~577V at 480 s for the frequencies of 100, 1000 and 2000 Hz, respectively. It is observed that after the initial stage, PEO under higher frequencies exhibits higher potential values. Meanwhile, louder acoustic emission was heard under the PEO at higher frequency. A higher cell potential may mean that thicker coatings are formed at higher frequencies. The negative cell potentials also show the initial potential surge, however, to much lower values of ~ 40 to 58 V, afterwards, they rise at different rates to the final voltages. The level of negative potential also increases with the increasing frequency. Fig.1(b) shows the cell potential-time responses under the unipolar regime. It is seen that the absence of negative potential has little influence on the positive cell potential, the curves are largely similar to those in the unipolar regime (Fig.1(a)).

The current waveforms at different frequencies were recorded and they are presented in Fig. 2. The waveform features a high spike of ~1.4 A/cm$^2$ and a subsequent plateau of ~1.1 A/cm$^2$ at 100 Hz, while the values of spike and plateau are ~2.5 A/ cm$^2$ and ~1.1 A/cm$^2$ at other two frequencies. According to Fig. 2, it is obvious that current with low frequency has a much longer pulse cycle compared with its counterparts, being 10, 1 and 0.5 ms for the three frequencies, indicating that the PEO discharging process is divided into shorter period at higher frequency (PEO is normally considered to happen at the positive pulse-on duration). Meanwhile, acoustic emission was heard to be sharper at higher frequencies.

Spark discharging is an important character of PEO, which can significantly influence the surface morphology and microstructure of the coating by the distribution and size of sparks. To explore the discharge difference among different frequencies (100 and 2000 Hz), appearances of sparks at various stages of PEO under bipolar regime were recorded and their images are presented in Fig. 3. The images in Fig.3 are recorded with the same exposure time of 0.125 ms, which is significantly less than the positive pulse-on time under 100 Hz and is comparable to the positive pulse-on time at 2000 Hz (0.1 ms). As compared in Fig. 3 (a) and (b), a large number of sparks existed on the sample surface at 100 Hz, while the number of sparks was smaller at 2000 Hz. The observed phenomenon in Fig.3 may indicate that less events of plasma discharge occurred at a fixed moment at the frequency of 2000 Hz. Although the size of sparks looks bigger at 100 Hz than that of 2000 Hz through the images taken by the digital camera in Fig.3, the real size of plasma sparks observed by naked eye appeared to be smaller at 100 Hz.

An interesting phenomenon was observed for the real time imaging at different frequencies, which is shown in Fig. 4. Fig.4(a) shows 14 images in sequential order taken from 300 s to 302 s for the PEO under 100 Hz. Among all images, only 4 images show the presence of plasma discharges on the sample surface. The detection of dark images is not unexpected since plasma discharge is normally thought to happen only at positive pulse-on time. It is clear in Fig.2 that the positive pulse only accounts for 1/5 duration in a cycle. Hence, theoretically, there is a probability of 4/5 for obtaining dark images, which coincide well with the result at 100 Hz. However, the case is quite different when the PEO frequency is 2000 Hz. Fig.4(b) shows that sparks are present in each of the images recorded under 2000 Hz in sequential order between 300 and 302 s. Images taken randomly at other PEO times, except those at the initial stage, also always show the presence of sparks. It is shown in Fig.2(c) that the pulse-off time and the negative pulse duration in a cycle is 0.4 ms, which is sufficiently longer than the camera exposure time of 0.125 ms. Hence, it is impossible that all the images at 2000 Hz in the present study were taken by accidental at the duration of positive pulse-on time. Then an interesting question arises: Why sparks appeared in those images, if they were taken at the pulse-off time of the 2000 Hz waveforms? The answer to this question is, we think, that unquenched discharge channels continues to give off light emission at the duration of pulse-off and negative pulse durations. After the initiation of a plasma discharge, the coating materials are melted within the discharge channels due to the passing of strong current density [39], and the discharge channels begin to quench after the cut-off of the anodic current at the negative and pulse-off times. However, due to the short pulse-off and negative pulse duration in a cycle at 2000 Hz, it is very likely that the molten discharge channels generated during the previous positive pulse duration were still kept at high temperatures with light emission even after the cut-off of anodic current. Hence, sparks are always detected in the images taken at 2000 Hz.

*3.2 Coating thickness, surface roughness, color and gas evaluation under different conditions*

Coating thickness was determined by eddy current method and is shown in Fig. 5. All the coatings in Fig.5 were formed under the bipolar regime for 480 s. It can be noted that coating fabricated at 2000 Hz is the thickest, reaching up to an average value of 67.4 ±4.2 μm. Coating thickness decreased along with decrement of frequency, being 52.9±3.1 μm at 100 Hz. The value of $R_a$ (arithmetical mean deviation of the roughness profile according to ISO-4287 standard) and $R_z$ (maximum height of profile) for different samples are also compared in Fig. 5. A similar trend to that of coating thickness can be noted, namely lower frequency resulted in a lower surface roughness while a higher frequency given rise to a much rougher surface. Similar trend was also reported in other studies [44,45]. This phenomenon was attributed to increment in diameter of discharge channels and height of pancakes structures [46].

Images of coatings fabricated with different frequencies under the bipolar and unipolar regimes respectively are presented in Fig. 6. The coatings are formed for 480 s. It can be noted that there is some difference between coatings formed at different frequencies. As indicated in Fig. 6 (a-c), at bipolar regime, coating formed at 100 Hz was dark gray in color, while it was black at 1000 Hz; when the frequency was further increased to 2000 Hz, the color transformed to dark black. The coatings formed without negative pulse show a similar trend, as illustrated in Fig. 6 (d-f). However, the coatings fabricated at 1000 and 2000 Hz under the unipolar regime seem to be darker than the corresponding coatings under the bipolar regimes.

A more accurate characterization of the colors of the samples has been adopted in this study. Fig. 7 shows the CIE L*a*b* color coordinate values of the PEO treated samples formed at various frequencies for 480 s. Color difference ($\triangle$E*) between samples and the correction whiteboard (L=73.94, a=16.32 and b=18.00) which was used as the reference of color measurement are also compared in Fig. 7. It seems that frequency had little influence on the values of a*, being -0.92±0.02, -0.91±0.04 and -0.99±0.04 for samples fabricated by bipolar regime at 100, 1000 and 2000 Hz respectively. As for the b*values, more negative values were obtained with increment of frequency, indicating a blue direction tendency [33]. All of the values of a* and b* for coatings obtained by unipolar regime were higher than that of coating fabricated through bipolar regime. Lightness of the coatings, represented by values of L*, being 65.19±0.16, 63.95±0.47 and 62.77±0.24 for three samples obtained at bipolar regime respectively, decreased with increment of frequency, this is in consistent with appearance of images presented in Fig.6. Brightness of coatings by unipolar regime were also lower than that of samples without application of negative pulse, being 64.30±0.03, 61.02± 0.04 and 60.73±0.06 respectively. As for △E*, since a whiteboard was chosen as the reference, hence the bigger value of △E* was, the deeper blackness the tested sample would be. It can be noted from Fig. 7(d) that the values of △E* increased with increment of frequency, being 29.90, 30.87 and 31.33 for three frequencies with negative pulse respectively, implying the higher the frequency was, the darker the coating was. At unipolar regime, the values of △E* between PEO coatings and the correction whiteboard were wider than that of bipolar regime, signifying that the absence of negative pulse could result in a darker film in the present study.

*3.3 Gas evaluation at different frequencies by bipolar regime*

Anomalous gas evolution is a characteristic of the PEO process, which is caused by the direct thermal decomposition of water within the discharge channels [39]. The volume of gas evolution at frequencies under the bipolar regime has been collected. Fig. 8 compared the influence of frequency on gas liberation behaviors at different stages of PEO process under various frequencies. At the first 180 seconds, volumes of gas liberated at different frequencies showed little distinction, after 300 seconds, the distinction was still not very evident, with gas volume collected at 1000 and 2000 Hz slightly larger than that of 100 Hz; however, with the reaction proceeded, the volume of gas evolved at higher frequency was obviously greater, being 234.7±1.7, 264.3±1.5 and 284.8±1.8 $cm^3$ for 100, 1000 and 2000 Hz after treated for 480 seconds respectively. This result is not out of expectation since the size of gas bubbles released at higher frequency observed by the naked eye was bigger compared with that of lower frequency at the later stage of PEO processes, and this is in accordance with acoustic emissions described before.

*3.4 Surface and cross-sectional morphologies and W content for the coatings formed at different conditions*

The surface morphologies of coatings formed at different frequencies by bipolar for 180 seconds are presented in Fig. 9. As shown in Fig. 9 (a), nodule-based structure randomly distributed on the surface for the coating formed at 100 Hz. No obvious micro pores were observed in the surface micrograph obtained at 100 Hz, only a few micro pores and other defects were observed in the cross section micrographs (Fig.9(b) and (c)). The surface of the coating formed at 2000 Hz was also featured by nodule-based structure (Fig.9(d)), however, a great number of micro pores are present on the coating surface, which is different from the morphology of the 100 Hz coating. The cross section of the coating formed at 2000 Hz also shows the presence of relatively big pores (see Fig. 9 (e) and (f)). It is generally considered that the micro pores are generated by the gas evolution associated with the intensive plasma discharges [38,39,47]. Hence, according to the coating morphology, the PEO at 2000 Hz might have been accompanied by heavier gas evolution and stronger discharges.

Fig. 10 illustrates the surface morphologies of coatings with longer treatment time of 480 seconds under bipolar regime at different frequencies. Typical PEO coatings with micro pores and micro cracks were observed for all the coated samples. The presence of micro cracks was attributed to thermal stress generated by the rapid solidification of molten products within the discharge channels [48]. Pancake structure was typical for all the coatings, as can be seen from the insets in Fig.10 (a), (b) and (c). Nevertheless, the difference between coatings was quite noticeable. There wasn't much protruding structure on the coating formed at 100 Hz, as indicated in Fig. 10 (a), signifying a smoother surface. With increment of frequency, the number of micro pores increased, especially for that of 2000 Hz. Furthermore, the size of micro pores increased along with increment of frequency. Meanwhile, there were a number of big micro holes on the coating formed at 2000 Hz; their diameter was larger than the usual micro pores. The cross section morphologies of PEO coatings by bipolar regime for 480 seconds are also presented in Fig. 10. The coating formed at 100 Hz exhibited a dense and uniform morphology except for a few small pores in the cross section. As the frequency was increased to 1000 and 2000 Hz, pores with very large dimensions (up to 50 μm in length) are observed at the coating cross section. The result in Fig.10 confirms that coatings with higher porosity are formed under higher frequencies, which is contrary to the works of Arunnellaiappan [27] and Kaseem et al [28].

The elemental compositions of the coatings formed at different frequencies as shown in Fig.10 were tested by EDS. The analyses were made on a large area on the coating surfaces and the results are summarized in Table 1. The EDS analyses show that O, Mg, Al and W are the major elements in the coating. Interestingly, the atom ratio of W in the coating increased notably with increment of frequency. It has been known that the incorporation of transition metal ions like Fe, Co, Ni, V and W favors development of black coatings [34,49]-51]. Obviously, the black color of the present coatings is associated with the incorporation of W species into the coatings. The darker appearance of coatings formed at high frequencies implies the incorporation of more W content in the coatings, which is consistent with the present EDS analyses.

Micrographs of coatings formed under unipolar regime for 480 seconds are illustrated in Fig. 11. Instead of forming the pancake structures as shown in Fig.10 (a), the surface of unipolar coating formed at 100 Hz was characterized by some open grooves and also the pancakes(see the inset). The surface of the unipolar coating formed at 2000 Hz is shown in Fig.11(b), which is featured by micro pores and pancake structures. Similar to the coatings formed under the bipolar regime, large pores are present at the cross section of the bipolar coating at 2000 Hz. The compositions of the coatings formed under unipolar regime have also been determined by EDS analyses made on the coating surfaces. Table 2 shows the results. Even higher W contents have been detected for the coatings formed without the negative pulse. The values of the atomic ratio of W to Al for the bipolar coatings are 0.08, 0.11 and 0.19 at 100, 1000 and 2000 Hz respectively, whilst they are 0.12, 0.24 and 0.26 for the corresponding unipolar coatings.

## 4. Discussion

It is obvious that the current frequency applied during PEO treatment decisively influenced the PEO process, resulting in coatings with higher thickness, roughness and porosity. Coatings obtained at 1000 and 2000 Hz are also incorporated with more W species which leads to a darker appearance. We found, after the observation of the microdischarge at different frequencies and the real time imaging results, that all these phenomena can be explained based on the influence of frequency on the microdischarge behavior during PEO.

According to the phenomenon observed during experiment and structures of coatings, a model describing the influence of frequency to the coating formation mechanism is proposed. Fig.12(a) and (b) illustrate the PEO under low frequency while those under higher frequency are depicted in Fig.12(c) and (d). It is shown in Fig.12 that the number of plasma discharges at high frequency is lower while their size is higher than that at low frequency. This fact is supported by the real time imaging results in this study. It has been mentioned previously that the quenching time for a microdischarge at high frequency is significantly shorter than that at low frequency, such as 100 Hz. Hence, the molten discharge channels at high frequency may have not been totally cooled down after the advent of next positive current pulse. The unquenched discharge channels likely possess a lower electric resistance, hence subsequent breakdown are prone to happen at those unquenched discharge channels. That may be the reason why the number of discharges is reduced at the PEO under higher frequency. At the same time, the plasma discharges at higher frequency may be more powerful due to the reduced number. A more powerful discharge may have caused the decomposition of higher amount of water in the discharge channels [39], hence higher volumes of gas have been collected at higher frequencies in the present study. Fig.12 (b) and (d) show how the plasma discharges influenced the coating microstructure in cross section. Details of the model in Fig12.(b) and (d) can be found in our previous paper[39]. It is shown that ionized gas occupies the volume below the coating outer layer and the inner molten oxide at a discharge location. After the extinguishment of the plasma discharge, the volume occupied by the ionized gas in Fig.12(b) and (d) will form the pores in coating cross section. It is obvious that big pores will be formed at higher frequency due to its bigger-sized plasma discharges.

The more powerful plasma discharges at higher frequency may also have resulted in the increased coating thickness, possibly due to their capability to decompose more coating materials from the electrolyte species. An increased coating thickness will in turn favor the forming of less and stronger discharges due to its increased coating resistance. The increased roughness of the coatings obtained from higher frequency is also easy to comprehend, since the powerful discharge can lead to larger pancake structures in the coating morphology [39]. The incorporation of more W species at higher frequencies may also be a result of the increased micro discharge energy which can decompose more $WO_4^-$ species from the electrolyte.

In this study, we have also investigated the influence of bipolar and unipolar regime on the PEO of AZ31 magnesium alloy. Compared to the bipolar coating, the unipolar coating formed under the same frequency shows the incorporation of more W content, thus a darker coating appearance. The function of negative pulse has been investigated in our previous study of the PEO of a zirconium alloy [38]. It was found that the application of negative pulse to the pulsed PEO waveform can facilitate the species transportation process due to the additional stirring to the electrolyte layer near the electrode surface by hydrogen evolution. In the present study, the species of Al and W are mainly in the form of $Al(OH)_4^-$ and $WO_4^{2-}$ [39]. Both of them can participate in the PEO reactions to form coatings. However, the molar concentration of Al species in the present electrolyte was 0.122 mol/l, while it was only 0.04 mol/l for that of W. Hence, there are more $Al(OH)_4^-$ species than $WO_4^{2-}$ in the electrolyte. The hydrogen stirring effect may have favored the transportation of a greater amount of Al species to form the coating. As a result, the W/Al ratio is lower for the bipolar coating.

**5. Conclusions**

Plasma electrolytic oxidation of AZ31 magnesium was carried out in aluminate-tungstate electrolyte under frequencies of 100, 1000 and 2000 Hz. The influence of negative pulse on PEO was also investigated. The main conclusions are as follows:

1. Increment of frequency resulted in coatings with higher thickness, higher roughness and darker appearance. The coatings formed under 100 Hz are compact, while big pores are present in the surface and cross section of the coatings formed at 1000 and 2000 Hz. Higher volumes of gas evolution were detected for the PEO under higher frequencies.

2. PEO under 2000 Hz shows a lower number of plasma discharges than that at 100 Hz, but the size of the discharge is greater. Dark images are usually recorded by real time imaging the PEO sparks at 100 Hz, which is due to the cut-off of positive current in the waveform. However, except for the initial stage, sparks are present in the image taken at any time during the PEO under 2000 Hz, which means the persisting of light emission from the hot discharge channels after the cut-off of anodic current. This phenomenon may be caused by the short pulse-off time at 2000 Hz, which is not sufficient to cool the discharge channels down to low temperature.

3. The black color of the coatings is caused by the incorporation of W species. PEO under

higher frequencies leads to a higher W/Al ratio in the resultant coating. The absence of negative pulse can also result in coatings with more W species.

4. The higher coating porosity and W incorporation can be attributed to the stronger plasma discharges under higher frequencies.

## Acknowledgement

Financial support from the National Natural Science Foundation of China (Grant Number 51671084) is gratefully acknowledged by the authors.

**Table Captions:**

Table 1. Composition of the bipolar coatings formed for 480 s in aluminate-tungstate electrolyte at different frequencies. The EDS results were obtained from a large area on the coating surface.

Table 2. Composition of the unipolar coatings formed for 480 s in aluminate-tungstate electrolyte at different frequencies. The EDS results were obtained from a large area on the coating surface.

**Figure Captions:**

Figure 1. Cell potential-time responses (absolute values are given to the negative potential) for the PEO of AZ31 magnesium alloy in aluminate-tungstate electrolyte at different frequencies with/without negative pulse (a) bipolar regime; (b) unipolar regime.

Figure 2. Current waveforms recorded for the PEO of AZ31 magnesium alloy in aluminate-tungstate electrolyte under bipolar regime at different frequencies: (a) 100 Hz; (b) 1000 Hz; (c) 2000 Hz.

Figure 3. Real-time images of discharging sparks recorded at different stages of PEO of AZ31 magnesium alloy under different frequencies: (a) 100 Hz; (b) 2000 Hz. The exposure time of the camera is 0.125 ms and the area depicted is 10 mm×20 mm.

Figure 4. Sequential real-time images recorded during the PEO of AZ31 magnesium alloy under different frequencies between 300 and 302 s: (a) 100 Hz; (b) 2000 Hz. The exposure time of the camera is 0.125 ms and the area depicted is 10 mm×20 mm.

Figure 5. The effect of frequency on the thickness, surface roughness ( $R_a$, $R_z$) of the coatings formed for 480 s in aluminate-tungstate electrolyte under bipolar regime.

Figure 6. Images for the appearance of coatings formed at 100, 1000 and 2000 Hz respectively under bipolar regime (a-c) and unipolar regime (d-f).

Figure 7. The plots of a*, b* and L* values (a-c) in the CIE L*a*b* color coordinate for PEO coatings formed for 480 s under different frequencies with/without the negative pulse. The a* and b* values indicate the location of color in the red-green and yellow-blue color coordinates, respectively. The L* value indicates the lightness of color. The color difference △E*(d) is the color difference between the PEO coatings and the white reference board.

Figure 8. The volume of gas evolution as a function of treatment time during PEO of the alloy under bipolar regime at different frequencies.

Figure 9. The SEM micrographs of surface and cross section of coatings formed for 180 s in aluminate-tungstate electrolyte by bipolar regime at 100 Hz (a-c) and 2000 Hz (d-f).

Figure 10. The SEM micrographs of surface and cross section of coatings formed after 480 s PEO treatment of AZ31 magnesium alloy in aluminate-tungstate electrolyte by bipolar regime at 100 Hz (a, b), 1000 Hz (c, d) and 2000 Hz (e, f).

Figure 11. The SEM micrographs of surface and cross section of coatings formed after 480 s PEO treatment of AZ31 magnesium alloy in aluminate-tungstate electrolyte by unipolar regime at 100 Hz (a, b) and 2000 Hz (c, d).

Figure 12. Schematic illustration of the coating formation processes at low frequency (a, b) and high frequency (c, d).

Table 1. Composition of the bipolar coatings formed for 480 s in aluminate-tungstate electrolyte at different frequencies. The EDS results were obtained from a large area on the coating surface.

| Frequency | Element composition in Wt% and At% (the values in brackets) and the ratio of W to Al | | | | |
|-----------|------------------|------------------|------------------|------------------|-------------|
|           | O                | Mg               | Al               | W                | W/Al        |
| 100       | 36.39 (53.08)    | 23.19 (22.26)    | 26.46 (22.89)    | 12.97 (01.77)    | 0.49 (0.08) |
| 1000      | 36.49 (56.24)    | 29.19 (28.08)    | 19.33 (16.75)    | 15.06 (01.92)    | 0.78 (0.11) |
| 2000      | 36.43 (53.25)    | 25.49 (25.86)    | 16.42 (15.00)    | 21.60 (02.90)    | 1.32 (0.19) |

Table 2. Composition of the unipolar coatings formed for 480 s in aluminate-tungstate electrolyte at different frequencies. The EDS results were obtained from a large area on the coating surface.

| Frequency | Element composition in Wt% and At% (the values in brackets) and the ratio of W to Al | | | | |
|---|---|---|---|---|---|
| | O | Mg | Al | W | W/Al |
| 100 | 33.57 (55.18) | 9.78 (10.58) | 31.42 (30.63) | 25.22 (03.61) | 0.80 (0.12) |
| 1000 | 30.33 (53.04) | 19.31 (22.22) | 19.29 (20.01) | 31.07 (04.73) | 1.61 (0.24) |
| 2000 | 27.97 (50.83) | 19.81 (23.69) | 18.73 (20.18) | 33.49 (05.30) | 1.79 (0.26) |

**Figure 1**

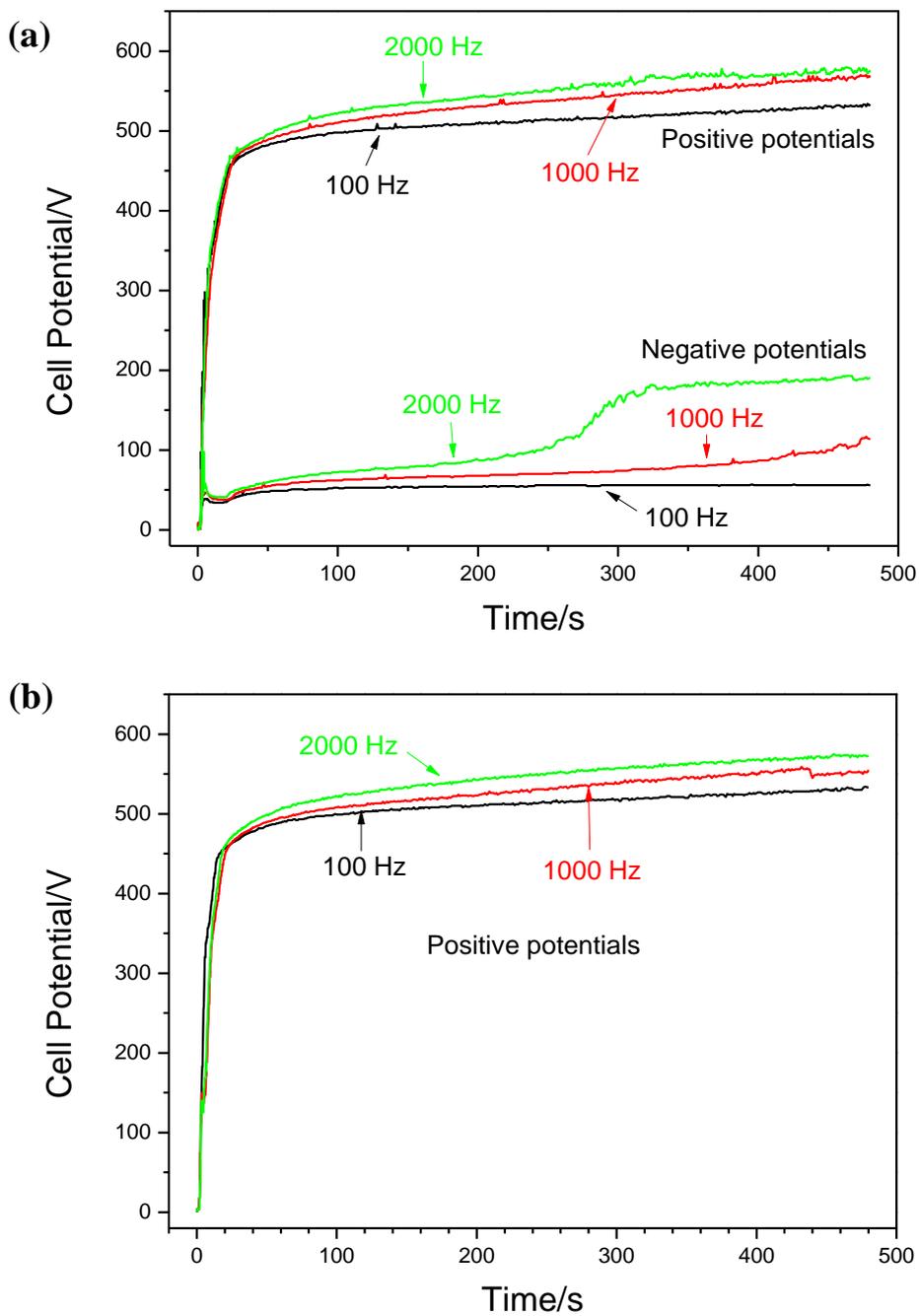

**Figure 2**

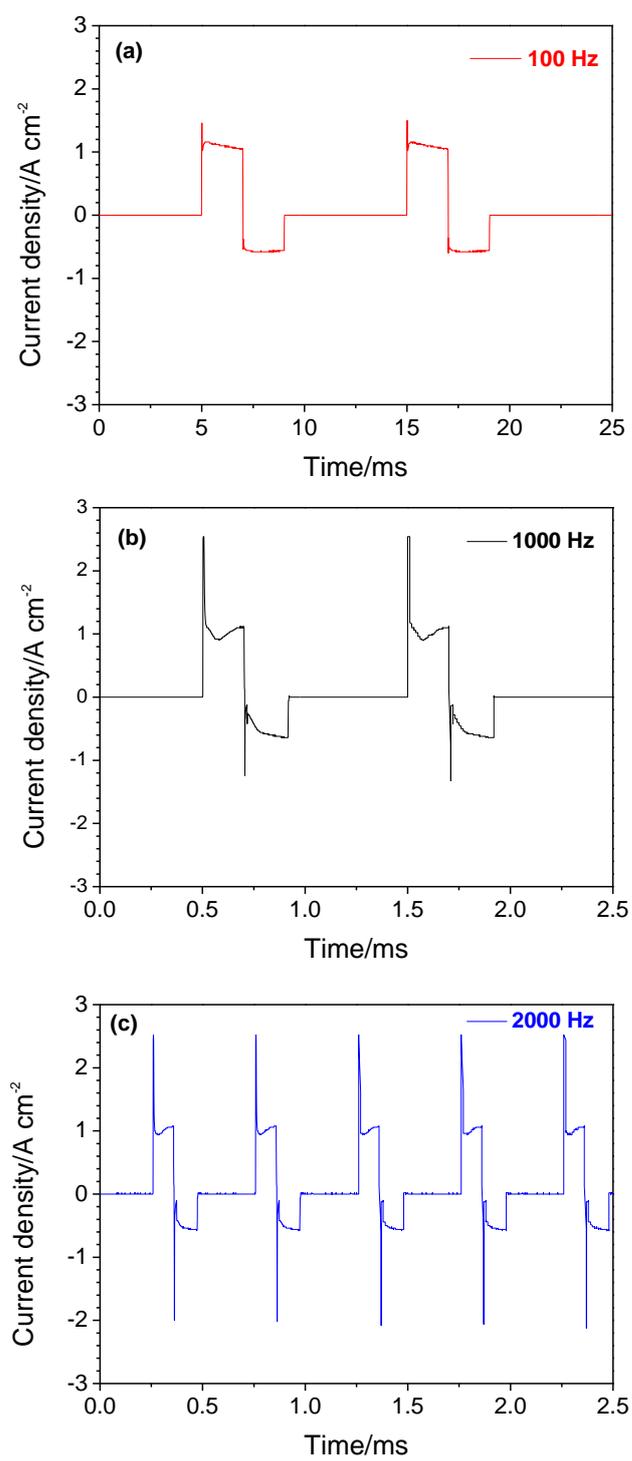

**Figure 3**

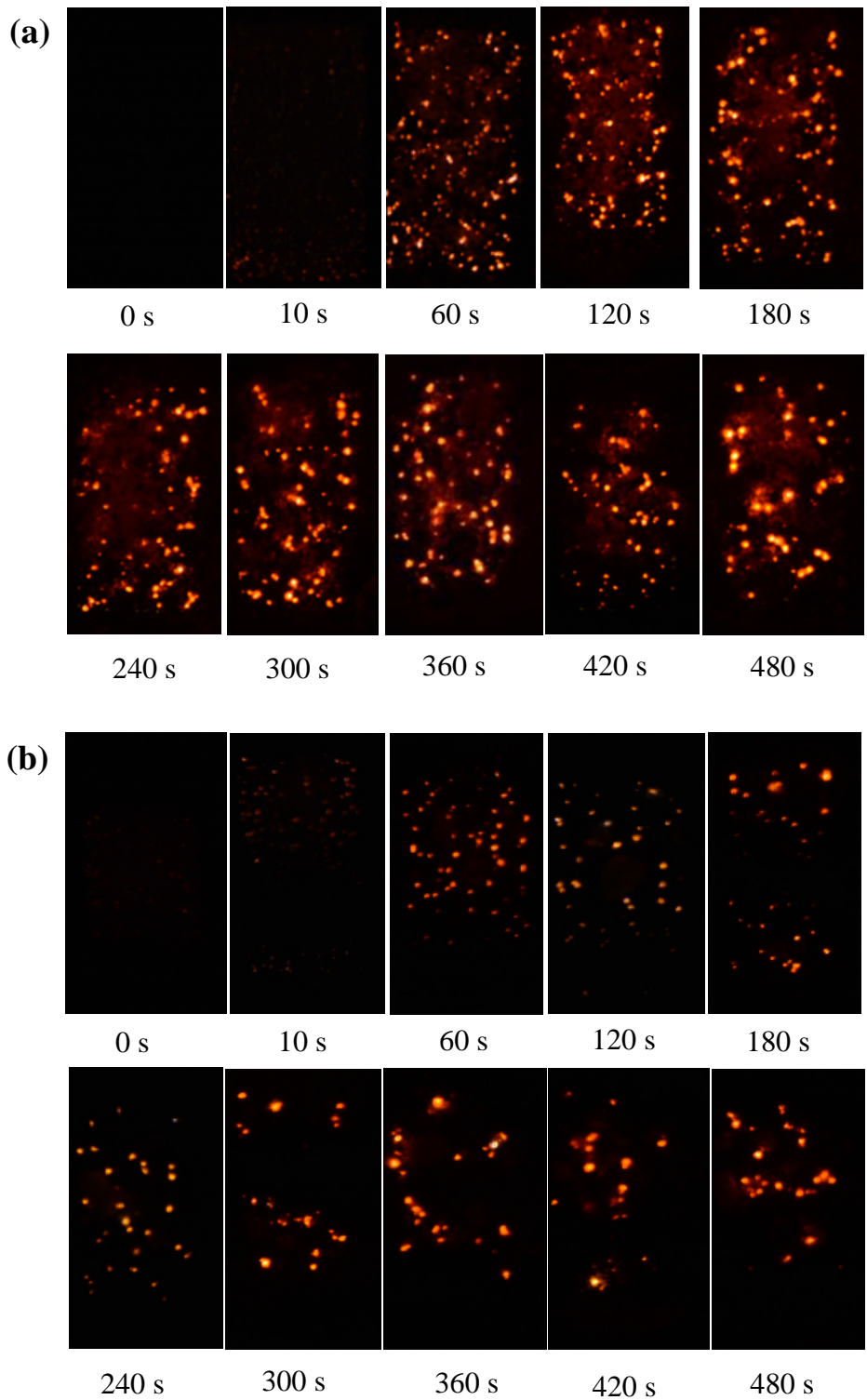

**Figure 4**

**(a)**

**(b)**

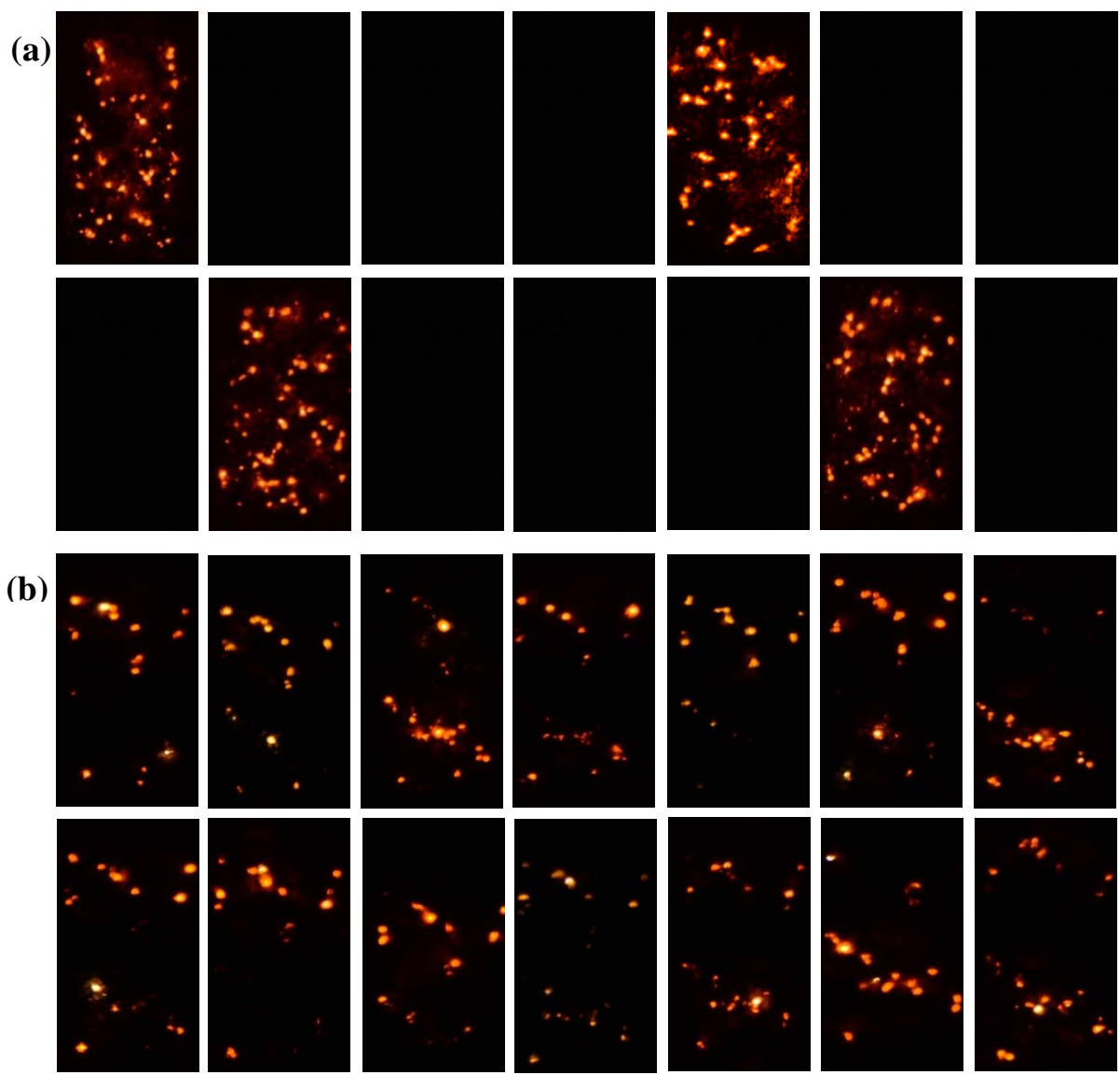

**Figure 5**

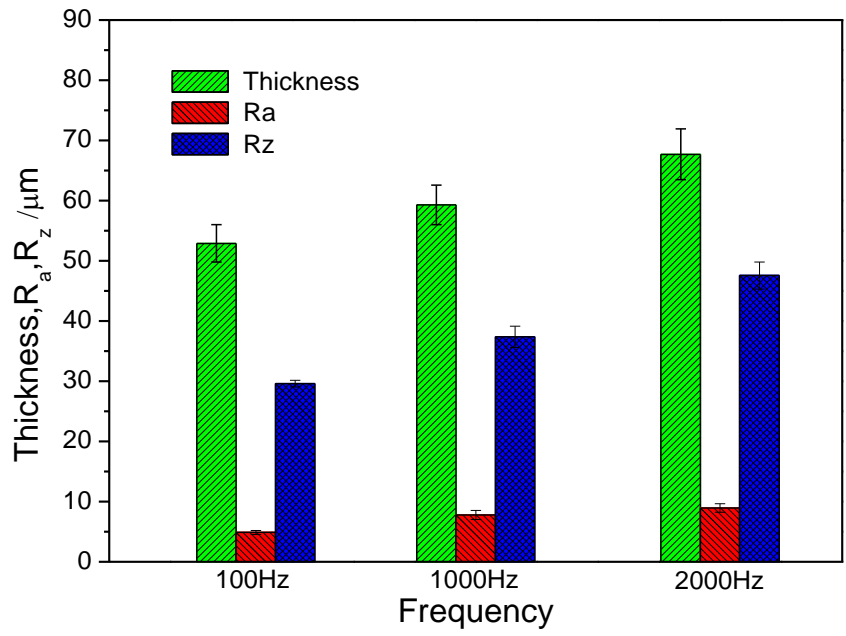

**Figure 6**

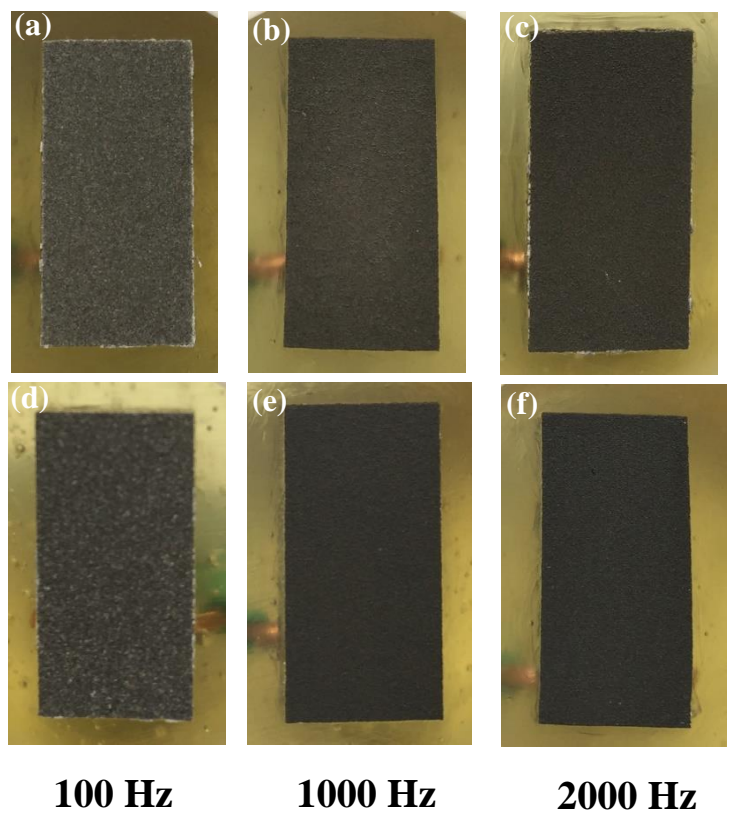

      **100 Hz**       **1000 Hz**     **2000 Hz**

**Figure 7**

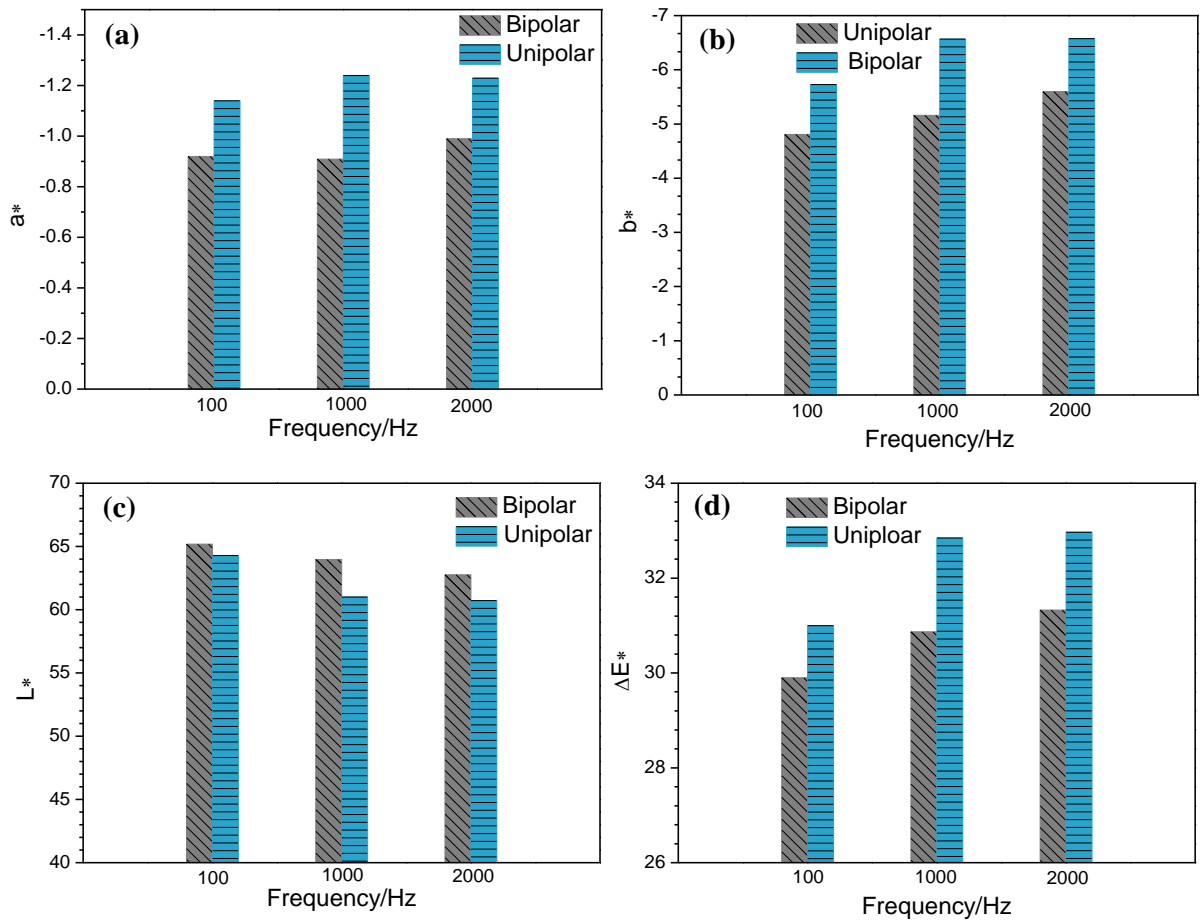

**Figure 8**

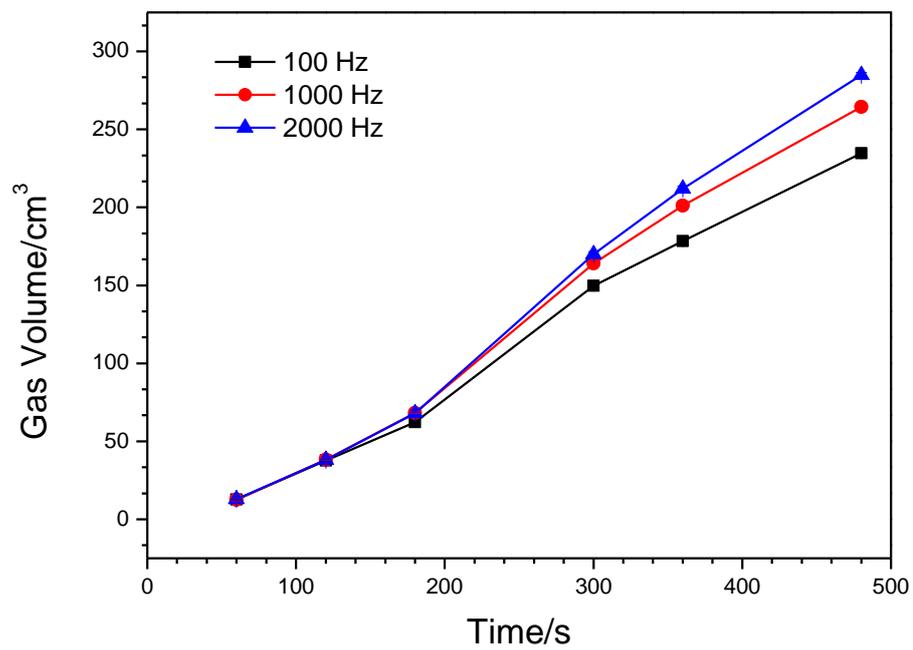

**Figure 9**

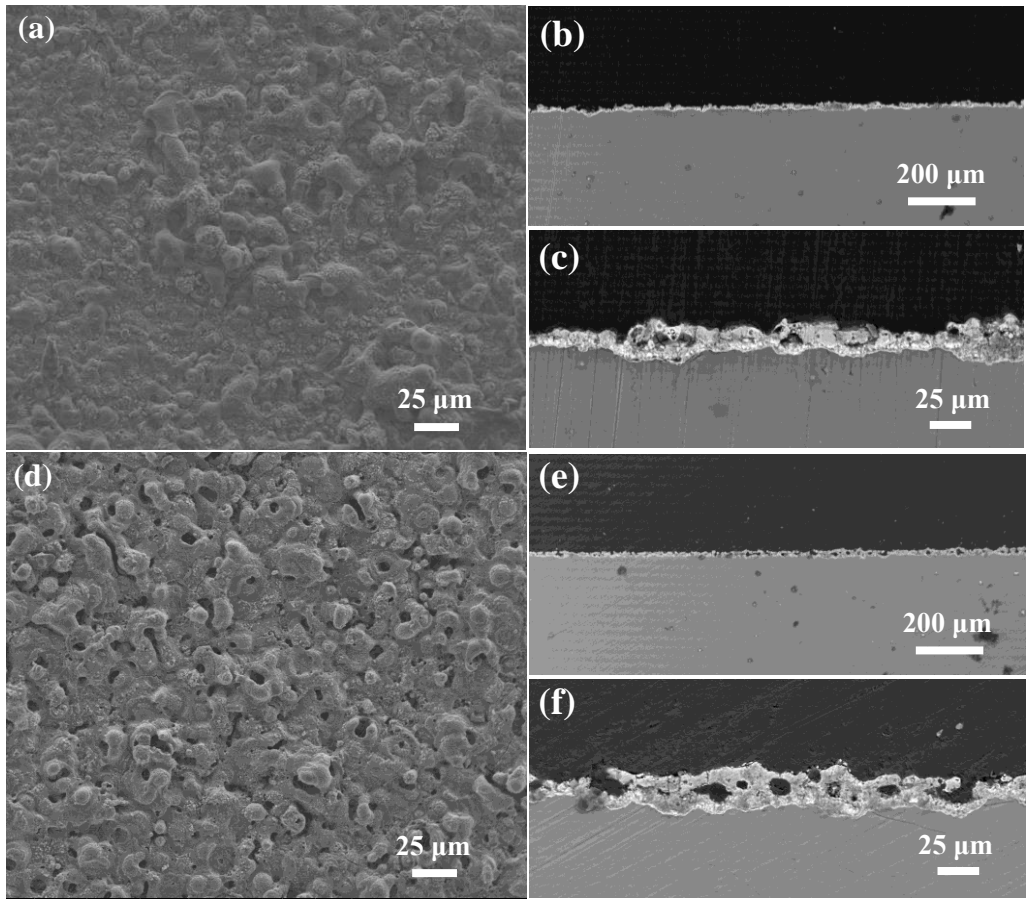

**Figure 10**

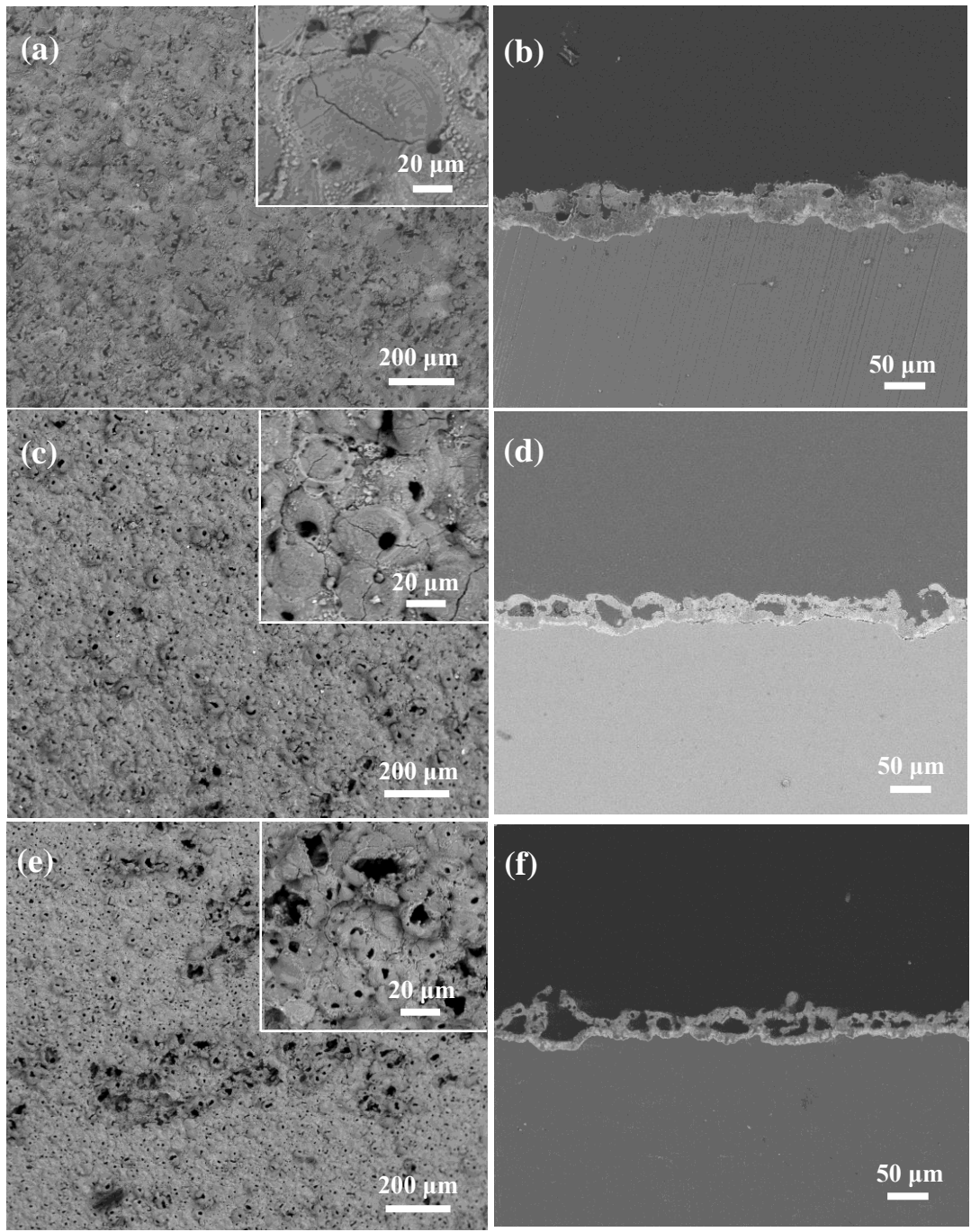

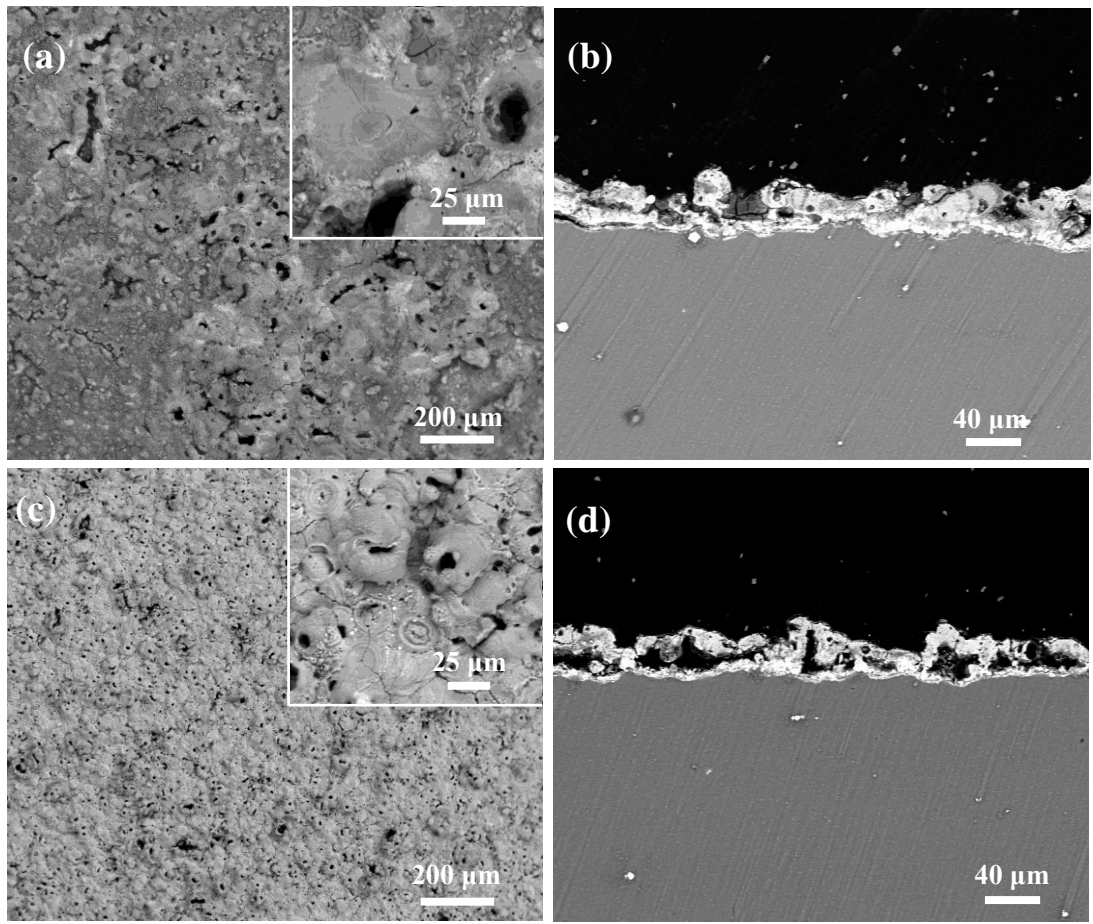

**Figure 12**

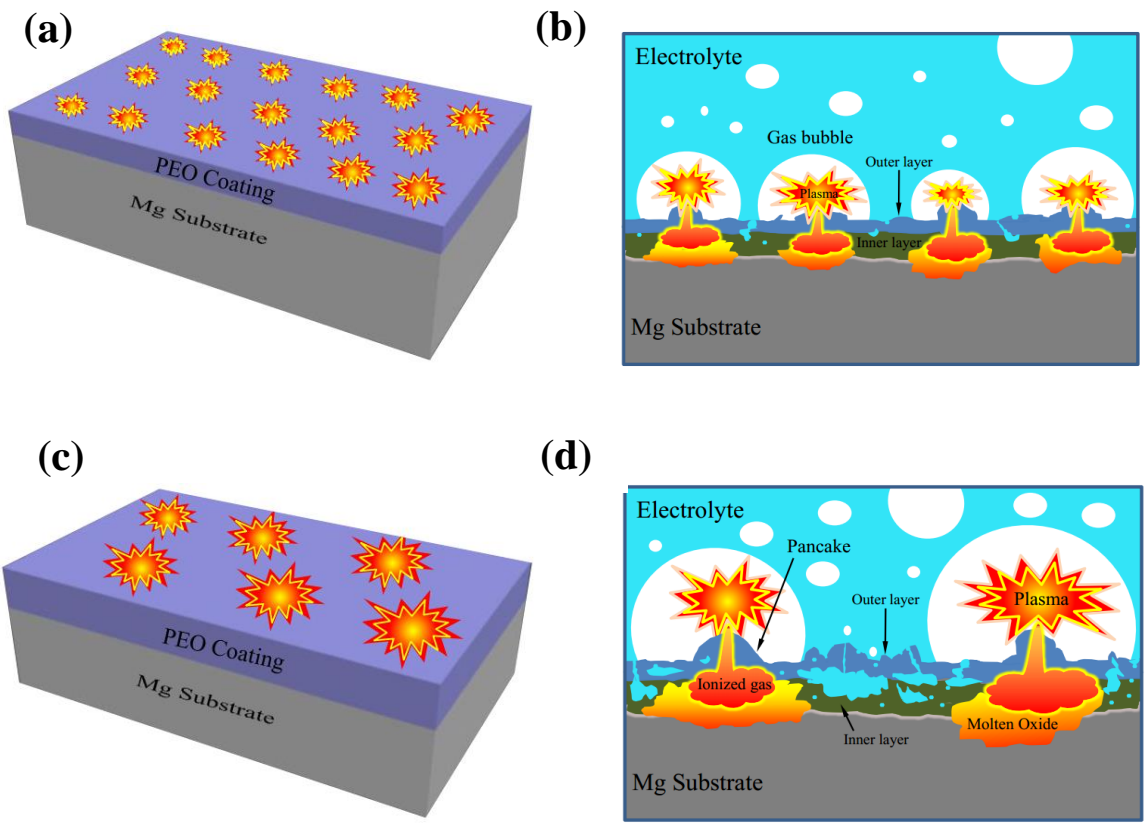

## Highlights

★ PEO was carried out at various current frequencies on AZ31 magnesium alloy.

★ High frequency results in darker coatings with higher porosity, thickness and roughness.

★ Light emission from hot discharge channels persists during the whole waveform cycle under high frequency.

★ The big pores are formed by the stronger plasma discharge at high frequency.

Graphical Abstract

**(a)**

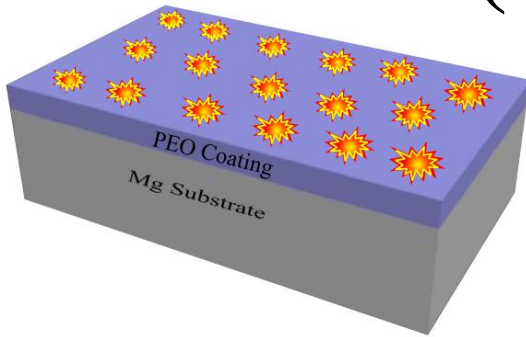 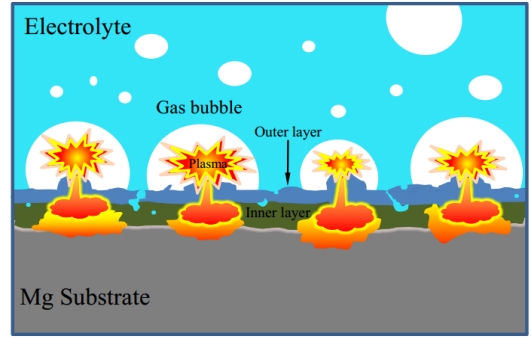

**(b)**

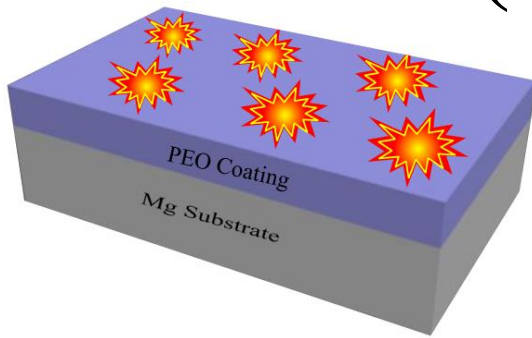 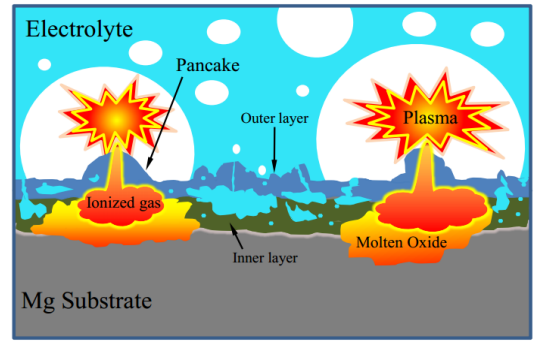

**(a) PEO under low frequency**     **(b) PEO under high frequency**